# M-PARTICLE QUANTUM WALKS WITH $\delta-$ INTERACTION

**CLEMENT AMPADU**

31 Carrolton Road
Boston, Massachusetts, 02132
U.S.A.
e-mail: drampadu@hotmail.com

**Abstract**

We consider directional correlations between $M-$particles on a line. For non-interacting particles we find analytic asymptotic expressions. When $\delta$ -interaction is introduced in the model we study the Fourier analysis and obtain general analytic formula for the wave function of the walk in the case $M=2$ for the transformation $C_{\delta}$, which can be considered an unfactorized version of the Hadamard walk in two-dimensions.

## 1. Introduction

Quantum walk on the line with a single particle possess a classical analogue, involving more walkers opens up the possibility to study collective quantum effect, such as many particle correlations. In this context, entangled initial states and indistinguishability of the particles play a role [4].

In this paper we investigate the non-classical effects in the $M-$particle discrete-time quantum walk on the line. Starting in Section 2 we recall some facts about the quantum walk of a single particle on a line, in Section 3 we extend the ideas of Section 2 to include $M$ particles, and give the probability $P_{sameside}(t)$ of finding the $M$ particles on the positive or negative side of the line. In Section 4 we obtain the asymptotic limit of $P_{sameside}(t)$. In Section 5 entangled initial states are considered, there we we consider two approaches in analyzing $P_{sameside}(t)$ for coin states that do not factorize. In Section 5.1 we analyze the case of maximally entangled Bell-type basis , and in Section 5.2 employ the equivalence between the $M$ -particle walk on the line and single particle walk on an $M-$dimensional lattice. This equivalence allows to invoke the weak limit theorem of Grimmet et.al [2] which have been successfully used by a number of authors [1,6]. In Section 6 we consider the probability $P_{sameside}(t)$ for indistinguishable particles, and show that for a particular choice of $M$ bosons or fermions the problem reduces to the case of distinguishable particles with maximally entangled coins.

Stefanak et.al [4] noted in their work involving directional correlations in quantum walk with two particles, that entanglement in two-particle non-interacting quantum walks cannot break the limit of probabilities they found for separable particles, and posed the following question: What happens if we consider interacting particles? This motivated them to introduce the concept of two-particle quantum walks with $\delta-$interaction to the solution of their question. The authors found out that by introducing a $\delta-$interaction one can exceed the limit derived for non-interacting particles. In Section 7 we commence the study of this new model focusing on the Fourier analysis. Section 8 is devoted to the conclusions.

**2. Quantum walk on the line with one particle**

The Hilbert space of the quantum walk is given by the tensor product $H = H_P \otimes H_C$, where $H_P$ is the position space of the particle , and $H_C$ is the coin space of the particle. The position space is spanned by the set of orthonormal states $\{|i\rangle : i \in Z\}$, and the coins space is spanned by $\{|L\rangle, |R\rangle\}$. To define the movement of the walker in one dimension, we first consider what happens on one step in the quantum walk. We first make superposition on the coin space with the coin operator $U_C$ and move the particle according to the coin state with the translation operator $S$ as follows $U_W = S \cdot (I \otimes U_C)$, where $I$ is the identity operator in the particle's position space, $U_W$ is the coin operator on the position space, and the translation operator $S$ is given by $S = \sum_x \{|x-1\rangle\langle x| \otimes |L\rangle\langle L| + |x+1\rangle\langle x| \otimes |R\rangle\langle R|\}$. The evolution of the quantum walk is then defined

by $\left|\Psi(t+1)\right\rangle = U_W \left|\Psi(t)\right\rangle$, which by induction on $t$, can be written in terms of the initial state of the particle as $\left|\Psi(t+1)\right\rangle = U^t{}_W \left|\Psi(0)\right\rangle$. If we take $U_C = H^*\left(\frac{1}{2},\frac{1}{2}\right)$, where $H^*(p,q)$ is a one-dimensional generalization of the Hadamard walk, for example, Ampadu [1], we get the most studied case of the Hadamard coin. We can express the wave vector $\left|\Psi(t)\right\rangle$ as the spinor $\left|\Psi(t)\right\rangle = \sum_{x\in Z}\begin{bmatrix} L_x(t) \\ R_x(t)\end{bmatrix}\left|x\right\rangle$, where the quibit $(L,R)^T$ has an upper (lower) component associated to the left (right) chirality, and the states $\left|x\right\rangle$ are the eigenstates of the position operator corresponding to the site $x$ on the line, from which it follows that probability distribution generated by the quantum walk is given by $p(x,t) = \left|L_x(t)\right|^2 + \left|R_x(t)\right|^2$.

**3. Quantum Walk on the line with M particles.**

We assume that the $M$ particles are distinguishable. Let $H_1,\ldots,H_M$ be the Hilbert spaces of the $M$ particles, and let $U_1,\ldots,U_M$ be the coin operator on the position space of the $M$ particles respectively, then the extension of the formalism above for an $M-$particle walk is as follows. For the $M$ distinguishable particles, the Hilbert space of the composite system can be written as $H^{composite} = H_1 \otimes H_2 \otimes H_3 \otimes \ldots \otimes H_{M-1} \otimes H_M$. Assuming further that the time evolution of the particles is independent, then the coin operator on the position space of the composite system can be written as $U^{composite} = U_1 \otimes U_2 \otimes U_3 \otimes \ldots \otimes U_{M-1} \otimes U_M$. We suppose further that the all the particles start from the origin then the initial state of the $M-$particle quantum walk is given by $\left|\Psi(0)\right\rangle = \left|\underbrace{0,0\ldots,0}_{M-times}\right\rangle \otimes \left|\Psi_C\right\rangle$, where $\left|\Psi_C\right\rangle$ is the initial coin state of the $M$ particles. Suppose that the initial coin state is factorizable. Since the time evolution of the particles are independent, entanglement is neglible and the factorization of the $M$-particle state remains unaffected, thus the joint probability distribution $p(m_1, m_2, \ldots, m_M, t)$ of finding the $M$ particles at sites $m_1, m_2, \ldots, m_M$, respectively at time $t$ can be written as the product of the single particle distribution to give

$p(m_1, m_2, \ldots, m_M, t) = \prod_{i=1}^{M} p(m_i, t)$, where $p(m_i, t)$ is the probability distribution of a single-particle quantum walk given that the initial coin state was $|\psi_i\rangle$. If the initial coin state is not factorizable, then $p(m_1, m_2, \ldots, m_M, t) \neq \prod_{i=1}^{M} p(m_i, t)$, however we can map the $M$-particle walk on a line to a quantum walk of a single particle on lattice whose dimension is the same as the number particles in the $M$-particle walk on the line, namely $M$ itself, then it follows that we can write coin operator on the position space of the composite system as $U^{composite} = S^{composite}\left(I^{composite} \otimes \left(\overbrace{U_C \otimes \ldots \otimes U_C}^{M-times}\right)\right)$, where

$I^{composite}$ is the identity operator on the position space of the composite system, and $S^{composite} = S_1 \otimes \ldots \otimes S_M$, and each $S_i$ in the tensor product represents the translation operator of the individual particles. Henceforth we will take $U_C = H^*\left(\frac{1}{2}, \frac{1}{2}\right)$, where $H^*(p,q)$ is a one-dimensional generalization of the Hadamard walk, for example, Ampadu [1], so

$U^{composite} = S^{composite}\left(I^{composite} \otimes \left(\overbrace{U_C \otimes \ldots \otimes U_C}^{M-times}\right)\right)$, the coin operator on the position space of the composite system can be interpreted as the coin operator of a single particle walk on an $M-$dimensional lattice with the coin given by the tensor product of $M$ Hadamard operators. The directional correlation between particles can be quantified in terms of the probability that the particles are found after $t$ steps of the quantum walk on the same side of the line Stefanak et.al [4]. In the case of the $M-$particle quantum walk, if all the particles are distinguishable, letting $P_{sameside}(t)$ denote the probability that the particles are found after $t$ steps of the quantum walk on the same side of the line, we have

$$P_{sameside}(t) = \sum_{m_1=-t}^{0} \sum_{m_2=-t}^{0} \cdots \sum_{m_{M-1}=-t}^{0} \sum_{m_M=-t}^{0} p(m_1, m_2, \ldots, m_M, t) + \sum_{m_1=1}^{t} \sum_{m_2=1}^{t} \cdots \sum_{m_{M-1}=1}^{t} \sum_{m_M=1}^{t} p(m_1, m_2, \ldots, m_M, t)$$

If the particles are indistinguishable, then the probabilities $p(m_1, m_2, \ldots, m_M, t)$ and $p(m_M, m_{M-1}, \ldots, m_1, t)$ are equivalent, hence $P_{sameside}(t)$ have to be restricted over an ordered pair $(m_1, m_2, \ldots, m_M)$ with $m_i \geq m_{i+1}$ for all $i = 1,2,3,\ldots, M-1$. In particular we have

$$P_{sameside}(t) = \sum_{i=1}^{M} \left[ \sum_{m_i+1=-t}^{0} \left( \sum_{m_i+1=m_i}^{0} p(m_1, m_2, \ldots, m_M, t) \right) + \sum_{m_{i+1}=1}^{t} \left( \sum_{m_i=m_i+1}^{t} p(m_1, m_2, \ldots, m_M, t) \right) \right].$$

**4. Separable initial states**

For the $M$ particles assuming the initial coin state $|\Psi_C\rangle$ can be factorized as $|\Psi_C\rangle = |\psi_1\rangle \otimes |\psi_2\rangle \otimes \ldots \otimes |\psi_M\rangle$, then we recall from the previous section that the joint probability distribution $p(m_1, m_2, \ldots, m_M, t)$ factorizes, so we can write the probability $P_{sameside}(t)$ as $P_{sameside}(t) = \prod_{i=1}^{M} P_i^-(t) + \prod_{i=1}^{M} P_i^+(t)$, where $P_i^{\pm}(t)$ denotes the probability that the particle has started the quantum walk with the coin state $|\psi_i\rangle$ is on the positive or negative side of the line after $t$ steps, in particular, $p_i^+(t) = \sum_{m_i=1}^{t} p(m_i, t)$ and $p_i^-(t) = \sum_{m_i=-t}^{0} p(m_i, t)$. Let us consider a general separable coin state of the form

$|\Psi_C\rangle = (a_1|L\rangle + b_1|R\rangle) \otimes \ldots \otimes (a_{M-1}|L\rangle + b_{M-1}|R\rangle) \otimes (a_M|L\rangle + b_M|R\rangle)$. The asymptotic distribution for a single particle due to Konno [3] can be written as, Stefanak et.al [4],

$$p(x,t,a_i,b_i) = \frac{1 - \frac{x}{t}\left((a_i + b_i)\bar{a}_i + (a_i - b_i)\bar{b}_i\right)}{\pi t \sqrt{1 - 2\frac{x^2}{t^2}\left(1 - \frac{x^2}{t^2}\right)}} \qquad (4.1)$$

The probability that the single particle is on the negative or positive side of the line has been calculated by Stefanak et.al [4] and are given as

$$p_i^-(a_i,b_i) = \int_{\frac{-t}{\sqrt{2}}}^{0} p(x,t,a_i,b_i)\,dx = \frac{1}{4}\left(2+\left((a_i+b_i)\bar{a}_i\right)+\left((a_i-b_i)\bar{b}_i\right)\right) \tag{4.2}$$

$$p_i^-(a_i,b_i) = \int_{0}^{\frac{t}{\sqrt{2}}} p(x,t,a_i,b_i)\,dx = \frac{1}{4}\left(2-\left((a_i+b_i)\bar{a}_i\right)+\left((a_i-b_i)\bar{b}_i\right)\right) \tag{4.3}$$

Now substituting (4.2) and (4.3) into $P_{sameside}(t) = \prod_{i=1}^{M} P_i^-(t) + \prod_{i=1}^{M} P_i^+(t)$ we get

$$P_{sameside}(t) = \frac{1}{4^M}\left[\prod_{i=1}^{M}\left(2+\left((a_i+b_i)\bar{a}_i\right)+\left((a_i-b_i)\bar{b}_i\right)\right)+\prod_{i=1}^{M}\left(2-\left((a_i+b_i)\bar{a}_i\right)+\left((a_i-b_i)\bar{b}_i\right)\right)\right] \tag{4.4}$$

Now we recast (4.4) in a suitable form that one can use for example to determine when the particles are most likely to be on the same side or opposite side of the half-axis by looking into the direction of maximal bias in the probability distribution. Now we consider the basis formed by the eigenstates of the Hadamard coin, Stefanak et.al [4], have shown that

$H^*\left(\frac{1}{2},\frac{1}{2}\right)\left|\chi^\pm\right\rangle = \pm\left|\chi^\pm\right\rangle$ has the following expression in the standard basis,

$$\left|\chi^\pm\right\rangle = \frac{\sqrt{2\pm\sqrt{2}}}{2}\left|L\right\rangle + \frac{\sqrt{2\mp\sqrt{2}}}{2}\left|R\right\rangle \tag{4.5}$$

Further they have shown that by decomposing the coin state of one particle in the Hadamard basis as $\left|\psi_i\right\rangle = h_i^+\left|\chi^+\right\rangle + h_i^-\left|\chi^-\right\rangle$, the relation between the coefficients in the standard and the Hadamard basis is given by $a_i = \frac{\sqrt{2+\sqrt{2}}}{2}h_i^+ + \frac{\sqrt{2-\sqrt{2}}}{2}h_i^-$ and $b_i = \frac{\sqrt{2-\sqrt{2}}}{2}h_i^+ + \frac{\sqrt{2+\sqrt{2}}}{2}h_i^-$, from which it follows that equation (4.4) can be written as

$$P_{sameside}(t) = \frac{1}{4^M}\left\{\prod_{i=1}^{M}\left[2+\sqrt{2}\left(\left|h_i^+\right|^2 - \left|h_i^-\right|^2\right)\right] + \prod_{i=1}^{M}\left[2-\sqrt{2}\left(\left|h_i^+\right|^2 - \left|h_i^-\right|^2\right)\right]\right\} \quad (4.6)$$

where the parameters $h_i^{\pm}$ are given by the by the overlap of the coin state $|\psi_i\rangle$ with the eigenstate $|\chi^{\pm}\rangle$.

## 5. Entangled initial states

### 5.1 On initial coin states of the Bell-type

If the coin state is not factorizable, then the joint probability distribution is no longer a product of the single-particle distribution. However, we can decompose the $M-$particle state in terms of single-particle amplitudes. In this way, we decompose the joint probability distribution into single-particle distributions plus an interefence term [4]. We then can use the results of the previous section to find the asymptotic value of the probability $P_{sameside}(t)$, say. We first consider the following *Bell-type Basis* for the $M-$particle quantum walk

$$|\psi^{\pm}\rangle = \frac{1}{\sqrt{2}}\left(|LR\ldots LR\rangle \pm |RL\ldots RL\rangle \pm |LR\ldots RL\rangle \pm |RL\ldots LR\rangle\right),\ |\phi^{\pm}\rangle = \frac{1}{\sqrt{2}}\left(|LL\ldots L\rangle \pm |RR\ldots R\rangle\right).$$

Note that we have used the term *Bell-type basis*, because the $M-$particle quantum walk is not a bipartite system, but the basis has the same form as the original Bell basis, and its self-explanatory for the $M-$particle quantum walk. Let $\psi_i^L(m_1,t)$ denote the amplitude of a single particle being after $t$ steps at the position $m_1$ with the coin state $|i\rangle,\ i=L,R$, provided that the initial coin state was $|L\rangle$ . Similarly, let $\psi_i^R(m_1,t)$ denote the amplitude of a single particle being after $t$ steps at the position $m_1$ with the coin state $|i\rangle,\ i=L,R$, provided that the initial coin state was $|R\rangle$. Let us agree to define

$$\psi_{k_1,k_2,\ldots,k_M}^{(LR\ldots LR)}(m_1,m_2,\ldots,m_M,t) \equiv \psi_{k_1}^L(m_1,t)\psi_{k_2}^R(m_2,t)\cdots\psi^L(m_{M-1},t)\psi^R(m_M,t) \quad (5.1)$$

$$\psi_{k_1,k_2,\ldots,k_M}^{(RL\ldots RL)}(m_1,m_2,\ldots,m_M,t) \equiv \psi_{k_1}^R(m_1,t)\psi_{k_2}^L(m_2,t)\cdots\psi^R(m_{M-1},t)\psi^L(m_M,t) \quad (5.2)$$

$$\psi^{(LR\ldots RL)}_{k_1,k_2,\ldots,k_M}(m_1,m_2,\ldots,m_M,t) \equiv \psi^{L}_{k_1}(m_1,t)\psi^{R}_{k_2}(m_2,t)\cdots\psi^{R}(m_{M-1},t)\psi^{L}(m_M,t) \tag{5.3}$$

$$\psi^{(RL\ldots LR)}_{k_1,k_2,\ldots,k_M}(m_1,m_2,\ldots,m_M,t) \equiv \psi^{R}_{k_1}(m_1,t)\psi^{L}_{k_2}(m_2,t)\cdots\psi^{L}(m_{M-1},t)\psi^{R}(m_M,t) \tag{5.4}$$

$$\psi^{(LL\ldots LL)}_{k_1,k_2,\ldots,k_M}(m_1,m_2,\ldots,m_M,t) \equiv \psi^{L}_{k_1}(m_1,t)\psi^{L}_{k_2}(m_2,t)\cdots\psi^{L}(m_{M-1},t)\psi^{L}(m_M,t) \tag{5.5}$$

$$\psi^{(RR\ldots RR)}_{k_1,k_2,\ldots,k_M}(m_1,m_2,\ldots,m_M,t) \equiv \psi^{R}_{k_1}(m_1,t)\psi^{R}_{k_2}(m_2,t)\cdots\psi^{R}(m_{M-1},t)\psi^{R}(m_M,t) \tag{5.6}$$

The joint probability distribution generated by the quantum walk of the $M-$particle with the initially entangled coins described by the Bell-type basis, upon using (5.1)-(5.6) is given by

$$p^{(\psi^{\pm})}(m_1,m_2,\ldots,m_M,t) = \frac{1}{2}\sum_{k_1,\ldots,k_M=L,R}\left|\begin{array}{l}\psi^{(LR\ldots LR)}_{k_1,k_2,\ldots,k_M}(m_1,m_2,\ldots,m_M,t) \pm \psi^{(RL\ldots RL)}_{k_1,k_2,\ldots,k_M}(m_1,m_2,\ldots,m_M,t) \pm \\ \psi^{(LR\ldots RL)}_{k_1,k_2,\ldots,k_M}(m_1,m_2,\ldots,m_M,t) \pm \psi^{(RL\ldots LR)}_{k_1,k_2,\ldots,k_M}(m_1,m_2,\ldots,m_M,t)\end{array}\right|^2$$

and

$$p^{(\phi^{\pm})}(m_1,m_2,\ldots,m_M,t) = \frac{1}{2}\sum_{k_1,\ldots,k_M=L,R}\left|\psi^{(LL\ldots LL)}_{k_1,k_2,\ldots,k_M}(m_1,m_2,\ldots,m_M,t) \pm \psi^{(RR\ldots RR)}_{k_1,k_2,\ldots,k_M}(m_1,m_2,\ldots,m_M,t)\right|^2$$

where the superscript in both the formulas $p^{(\psi^{\pm})}(m_1,m_2,\ldots,m_M,t)$ and $p^{(\phi^{\pm})}(m_1,m_2,\ldots,m_M,t)$ indicates the initial coin state. Since $H^{*}\left(\frac{1}{2},\frac{1}{2}\right)$ and the Bell-type initial states contain only real entries, we can drop the absolute value in the formulas for $p^{(\psi^{\pm})}(m_1,m_2,\ldots,m_M,t)$ and $p^{(\phi^{\pm})}(m_1,m_2,\ldots,m_M,t)$, and expand the joint probability distribution in the form

$$p^{(\psi\pm)}(m_1,m_2,\ldots,m_M,t) = \frac{1}{2}\left[\begin{array}{l}p^{LR\ldots LR}(m_1,m_2,\ldots,m_M,t) + p^{RL\ldots RL}(m_1,m_2,\ldots,m_M,t) + \\ p^{LR\ldots RL}(m_1,m_2,\ldots,m_M,t) + p^{RL\ldots LR}(m_1,m_2,\ldots,m_M,t)\end{array}\right] \pm \prod_{i=1}^{M}\varphi(m_i,t) \tag{5.7}$$

and

$$p^{(\phi\pm)}(m_1,m_2,\ldots,m_M,t) = \frac{1}{2}\left[p^{LL\ldots LL}(m_1,m_2,\ldots,m_M,t) + p^{LL\ldots LL}(m_1,m_2,\ldots,m_M,t)\right] \pm \prod_{i=1}^{M}\varphi(m_i,t) \tag{5.8}$$

Note that in (5.7) and (5.8) terms of the form $p^{LR\ldots LR}(m_1,m_2,\ldots,m_M,t)$, for example, are defined as

$$p^{LR\ldots LR}(m_1,m_2,\ldots,m_M,t)\equiv p^L(m_1,t)p^R(m_2,t)\ldots p^L(m_{M-1},t)p^R(m_M,t) \qquad (5.9)$$

and the terms in the product $\prod_{i=1}^{M}\varphi(m_i,t)$ are defined for example in (5.8) for $\varphi(m_1,t)$ as

$\varphi(m_1,t)\equiv\varphi_L^{LR\ldots LR}(m_1,t)+\varphi_R{}^{LR\ldots LR}(m_1,t)$ , where in this definition $\varphi_L^{LR\ldots LR}(m_1,t)$ for example is understood as $\varphi_L^{LR\ldots LR}(m_1,t)\equiv\varphi_L^L(m_1,t)\varphi_L^R(m_1,t)\ldots\varphi_L^L(m_{M-1},t)\varphi_L^R(m_M,t)$.

Recall for distinguishable particles,

$$P_{sameside}(t)=\sum_{m_1=-t}^{0}\sum_{m_2=-t}^{0}\cdots\sum_{m_{M-1}=-t}^{0}\sum_{m_M=-t}^{0}p(m_1,m_2,\ldots,m_M,t)+\sum_{m_1=1}^{t}\sum_{m_2=1}^{t}\cdots\sum_{m_{M-1}=1}^{t}\sum_{m_M=1}^{t}p(m_1,m_2,\ldots,m_M,t) \qquad (5.10)$$

If we insert either (5.7) and (5.8) into (5.10) we find we can write the result as

$p_{samside}^{\psi\pm}(t)=p_{sameside}^{LR\ldots LR}(t)\pm I(t)$ and $p_{samside}^{\phi\pm}(t)=p_{sameside}^{LL\ldots LL}(t)\pm I(t)$, where $I(t)$ is in the interference term, and is given by $I(t)=\left(\varphi^-(t)\right)^2+\left(\varphi^+(t)\right)^2$, where $\varphi^-(t)=\sum_{m_1=-t}^{0}\varphi(m_1,t)$ and

$\varphi^+(t)=\sum_{m_1=1}^{t}\varphi(m_1,t)$.

### 5.2 On general initial coin states

On the $M-$dimensional lattice the time evolution of the Hadamard walk is determined by the generator $U(k_1,k_2,\ldots,k_M)=U(k_1)\otimes\ldots\otimes U(k_{M-1})\otimes U(k_M)$, where $U(k_r)$ denotes the generator of a single-particle walk on the line and this is given by $U(k_r)=\left[D(e^{-ik},e^{ik})\cdot H^*\left(\frac{1}{2},\frac{1}{2}\right)\right]$. Due to the tensor product decomposition of $U(k_1,k_2,\ldots,k_M)$ we can write its eigenvalues as the *product* of the eigenvalues $U(k_r)$ and we can also write the eigenvectors of $U(k_1,k_2,\ldots,k_M)$ as the tensor product of the eigenvectors of $U(k_r)$. The eigenvalues of $U(k_1,k_2,\ldots,k_M)$ are given by

$\lambda_{x_1x_2\ldots x_M}(k_1,k_2,k_3,\ldots,k_M)=e^{i\left[\sum_{t=1}^{M}\left(w_{x_t}(k_t)\right)\right]}$, $x_1,x_2,\ldots,x_M=1,2$, where $e^{iw_{x_1}(k)}$ are the eigenvalues of the

matrix $U(k_r)$, and $w_{x_1}(k)$ is determined by $w_1(k) = \arcsin\left(\frac{\sin(k)}{\sqrt{2}}\right)$, $w_2(k) = \pi - w_1(k)$. Similarly, the eigenvectors of $U(k_1, k_2, \ldots, k_M)$ are given by the tensor product $v_{x_1 x_2 \ldots x_M}(k_1, k_2, \ldots, k_M) = v_{x_1}(k_1) \otimes v_{x_2}(k_2) \otimes \ldots \otimes v_{x_M}(k_M)$ of the eigenvectors of the matrices $U(k_r)$ which are given by $v_1(k) = \frac{1}{\sqrt{n_1(k)}}\left(e^{ik}, \sqrt{2}e^{iw_1(k)} - e^{ik}\right)^T$, $v_2(k) = \frac{1}{\sqrt{n_2(k)}}\left(-e^{ik}, \sqrt{2}e^{-iw_1(k)} + e^{ik}\right)^T$, where $n_1(k) = 2\left(1 + \cos^2(k) - \cos k\sqrt{1+\cos^2 k}\right)$ and $n_2(k) = 2\left(1 + \cos^2(k) + \cos k\sqrt{1+\cos^2 k}\right)$. Now invoking the weak limit theorem of Grimmett et.al, the cumulative distribution function is then given by

$$F(\tilde{t}_1, \tilde{t}_2, \tilde{t}_3, \ldots, \tilde{t}_M) = \sum_{x_1, \ldots, x_M = 1}^{2} \int_{\nabla w^{-1}{}_{x_1, \ldots, x_M}(I)} d\mu_{x_1 x_2 x_3 \ldots x_{M-1} x_M}$$

, where we have let $\tilde{t}_i = \frac{t_i}{t}$, and

$I = \left(-\infty, \tilde{t}_1\right) \times \left(-\infty, \tilde{t}_2\right) \times \cdots \times \left(-\infty, \tilde{t}_M\right)$. The probability measure is determined by

$$\mu_{x_1 x_2 x_3 \cdots x_M} = \left|\left(v_{x_1 x_2 x_3 \cdots x_M}(k_1, k_2, \cdots, k_M), \psi_C\right)\right|^2 \prod_{z=1}^{M}\left(\frac{dk_z}{2\pi}\right)$$

. Note that the vector $\psi_C$ which corresponds to the initial state of the coin $|\psi_C\rangle$ consists of $2^M$ – components. From the explicit form of the eigenvectors $v_{x_1 x_2 \ldots x_M}(k_1, k_2, \ldots, k_M)$, the probability measure $\mu_{x_1 x_2 x_3 \cdots x_M}$ can be seen to equal

$$\mu_{x_1 x_2 x_3 \cdots x_M} = \frac{1}{2^M}\left[1 + \sum_{i=1}^{M}(-1)^{x_i+1}\left(C_i C(k_i) + S_i S(k_i)\right) + \sum_{j=2}^{M}\left[(-1)^{\sum_{i=1}^{j} x_i}\left(C_{x_1 x_2 \cdots x_j}\prod_{i=1}^{j} C(k_i) + S_{x_1 \cdots x_j}\prod_{i=1}^{j} S(k_i) + \sum_{i=1}^{j} x_i C(k_i) \prod_{\substack{v=1 \\ v \neq i}}^{j} S(k_v)\right)\right]\right]\prod_{i=1}^{M}\left(\frac{dk_i}{2\pi}\right)$$

where $C(k) = \frac{\cos(k)}{\sqrt{1+\cos^2 k}}$, $S(k) = \frac{\sin(k)}{\sqrt{1+\cos^2 k}}$, and the coefficients $C, S, X$ are determined from the initial state of the coin $|\psi_C\rangle$. To obtain the cumulative distribution function we have need the domain of integration, and these are obtained from $\nabla w_{x_1 x_2 x_3 \ldots x_M}(k_1, k_2, \cdots, k_M)$, that is, the gradients of the phases $w_{x_1 x_2 x_3 \ldots x_M}(k_1, k_2, \cdots, k_M)$ of the eigenvalues of $U(k_1, \ldots, k_M)$. From the explicit form of the eigenvalues of $U(k_1, \ldots, k_M)$, the gradients are

$\nabla w_{x_1 x_2 x_3 \ldots x_M}(k_1, k_2, \cdots, k_M) = \left((-1)^{x_1+1} C_{x_1}, (-1)^{x_2+1} C_{x_2}, \cdots, (-1)^{x_M+1} C_{x_M}\right)$. Using the above results and the substitution $C(k_i) = \dfrac{\cos k_i}{\sqrt{1+\cos^2 k_i}} = q_i$, $dk_i = \dfrac{dq_i}{\left(1-q_i^2\right)\sqrt{1-2q_i^2}}$, the cumulative distribution function takes the form

$$F(\tilde{t}_1, \tilde{t}_2, \ldots, \tilde{t}_M) = \frac{1}{\pi^M} \prod_{i=1}^{M} \int_{-1\sqrt{2}}^{t_i} \frac{dq_i}{\left(1-q_i^2\right)\sqrt{1-2q_i^2}} \left[1 - \sum_{i=1}^{M} C_i q_i + \sum_{j=2}^{M} C_{x_1 x_2 \ldots x_j} \prod_{i=1}^{j} q_i \right].$$ Recall that

differentiation of the cumulative distribution function gives the probability density function, in particular the relation $p(g_1, g_2, g_3, \ldots, g_M) = \dfrac{\partial^M F}{\partial g_1 \partial g_2 \cdots \partial g_M}$, implies that

$$p(t_1, t_2, t_3, \ldots, t_M) = \frac{1}{\pi^M \prod_{i=1}^{M} \left[\left(1 - \frac{t_i^2}{t^2}\right)\sqrt{1 - \frac{2t_i^2}{t^2}}\right]} \left[1 - \sum_{i=1}^{M} \frac{C_i t_i}{t} + \sum_{j=2}^{M} \left( C_{x_1 x_2 x_3 \cdots x_j} \prod_{i=1}^{j} \frac{t_i}{t^j} \right)\right],$$ where we have

let $q_i = \dfrac{t_i}{t}$. Now using $p(t_1, t_2, t_3, \ldots, t_M)$ and replacing the sums in

$$P_{sameside}(t) = \sum_{m_1=-t}^{0} \sum_{m_2=-t}^{0} \cdots \sum_{m_{M-1}=-t}^{0} \sum_{m_M=-t}^{0} p(m_1, m_2, \ldots, m_M, t) + \sum_{m_1=1}^{t} \sum_{m_2=1}^{t} \cdots \sum_{m_{M-1}=1}^{t} \sum_{m_M=1}^{t} p(m_1, m_2, \ldots, m_M, t)$$ by

integrals we get

$$P_{sameside}(t) = \prod_{i=1}^{M} \left[ \int_{\frac{-t}{\sqrt{2}}}^{0} p(t_1, t_2, \ldots, t_M, t)\, dt_i \right] + \prod_{i=1}^{M} \left[ \int_{0}^{\frac{t}{\sqrt{2}}} p(t_1, t_2, \ldots, t_M, t)\, dt_i \right],$$ where $p(t_1, t_2, t_3, \ldots, t_M)$ is

given as above. Whilst we have not been able to obtain a closed form formula, we should note that there is a simplification of the formula above which will give the same formula in the case of seperable states, that is, for seperable states

$$P_{sameside}(t)=\prod_{i=1}^{M}\left[\int_{\frac{-t}{\sqrt{2}}}^{0} p(t_1,t_2,\ldots,t_M,t)\,dt_i\right]+\prod_{i=1}^{M}\left[\int_{0}^{\frac{t}{\sqrt{2}}} p(t_1,t_2,\ldots,t_M,t)\,dt_i\right]$$

$$=\frac{1}{4^M}\left\{\prod_{i=1}^{M}\left[2+\sqrt{2}\left(\left|h_i^+\right|^2-\left|h_i^-\right|^2\right)\right]+\prod_{i=1}^{M}\left[2-\sqrt{2}\left(\left|h_i^+\right|^2-\left|h_i^-\right|^2\right)\right]\right\}$$

This implies that the bounds obtained by both formulas is the same.

**6. Asymptotic Probability for Indistinguishable particles.**

It is natural in this case to use the so-called second quantization formalism [5].

To give the second quantization formalism for the single particle dynamics. Let us denote the bosonic creation operators by ${}^{T}\bar{\hat{a}}_{(m_1,r_1)}$ , and the fermionic creation operators similarly. Note that ${}^{T}\bar{\hat{a}}_{(m_1,r_1)}$ creates one bosonic particle at position $m_1$ with the internal state $\left|r_1\right\rangle$, $r_1 = L, R$; similarly for the fermionic operator. The dynamics of the quantum walk with indistinguishable particles on a one particle level, for bosonic particles is given by the following transformation of the creation operators (which is also similar for fermionic particles)

$${}^{T}\bar{\hat{a}}_{(m_1,L)} \rightarrow \frac{1}{\sqrt{2}}\left({}^{T}\bar{\hat{a}}_{(m_1-1,L)} + {}^{T}\bar{\hat{a}}_{(m_1+1,R)}\right)$$

$${}^{T}\bar{\hat{a}}_{(m_1,R)} \rightarrow \frac{1}{\sqrt{2}}\left({}^{T}\bar{\hat{a}}_{(m_1-1,L)} - {}^{T}\bar{\hat{a}}_{(m_1+1,R)}\right)$$

The difference between the bosonic and fermionic operators is that the bosonic operators fulfill the following commutation relations

$\left[\hat{a}_{(m_1,r_1)}, \hat{a}_{(m_1,r_2)}\right]=0$ and $\left[\hat{a}_{(m_1,r_1)}, {}^{T}\bar{\hat{a}}_{(m_1,r_2)}\right]=\delta_{m_1m_2}\delta_{r_1r_2}$

whilst the fermionic operators satisfy the anti-commutor relations

$\left\{\hat{b}_{(m_1,r_1)}, \hat{b}_{(m_2,r_2)}\right\}=0$ and $\left\{\hat{b}_{(m_1,r_1)}, {}^{T}\bar{\hat{b}}_{(m_1,r_2)}\right\}=\delta_{m_1m_2}\delta_{r_1r_2}$

Notice that since the dynamics is defined for a single particle, we can describe the state of the $M-$indistinguishable particles after $t$ steps of the quantum walk in terms of the single-particle amplitudes. As the initial state of the coin we choose $\left|\Psi(0)\right\rangle = \left|1_{0\cdots 0,LR\cdots LR}\right\rangle$, that is, the $M-$particles are initially at the origin with alternating coin states starting with $\left|L\right\rangle$ for the first particle, $\left|R\right\rangle$ for the second particle, and so on. Now recalling the amplitudes $\psi_i^L$ and $\psi_i^R$ for the single particle performing the quantum walk with the initial coin state $\left|L\right\rangle$, $\left|R\right\rangle$ respectively, the state of the $M$ bosons (and similarly for fermions) can be stated in the following form

$\left|\Psi^{Bosons}(t)\right\rangle = \sum_{m_1,\cdots,m_M} \sum_{r_1,\ldots,r_M=L,R} \psi_{r_1\cdots r_M}^{(LR\cdots LR)}(m_1,\cdots,m_M,t)^T\,\bar{\hat{a}}_{(m_1,\cdots m_M;r_1,\cdots r_M)}\left|vac\right\rangle$, where $\left|vac\right\rangle$ denotes the vacuum state, and we have defined $\psi_{r_1\cdots r_M}^{(LR\cdots LR)}(m_1,\cdots,m_M,t) \equiv \psi_{r_1}^L(m_1,t)\cdots\psi_{r_{M-1}}^L(m_{M-1},t)\psi_{r_M}^R(m_M,t)$ and similarly for ${}^T\bar{\hat{a}}_{(m_1,\cdots m_M;r_1,\cdots r_M)}$. Note that the summation indices $m_1,\ldots,m_M$ run over all possible sites.

Now we consider the joint probability $p(m_1,m_2,m_3,\cdots,m_M;t)$ that after $t$ steps we detect a particle at sites $m_1,m_2,m_{3,}\cdots,m_M$ simultaneously with $m_i \geq m_{i+1}$. If $m_i \neq m_{i+1}$, then,

$$p^{(Bosons,Fermions)}(m_1,m_2,m_3,\cdots,m_M;t) = \left|\left\langle 1_{(m_1,m_2,\ldots,m_M;r_1,r_2,\ldots r_M)}\right|\Psi^{(Bosons,Fermions}(t)\rangle\right|$$

$$= \sum_{r_1,\cdots r_M=L,R}\left|\Psi_{r_1\cdots r_M}^{LR\cdots LR}(m_1,m_2,\cdots,m_M,t) \pm \Psi_{r_1\cdots r_M}^{RL\cdots RL}(m_1,m_2,\cdots,m_M,t) \pm \Psi_{r_1\cdots r_M}^{LR\cdots RL}(m_1,m_2,\cdots,m_M,t) \pm \Psi_{r_1\cdots r_M}^{RL\cdots LR}(m_1,m_2,\cdots,m_M,t)\right|^2$$

where the plus sign on the right corresponds to bosonic, and the negative sign corresponds to fermionic. If we compare these expressions to those for the $Bell-type$ states considered previously we get the following relations $p^{Bosons}(m_1,m_2,\cdots,m_M;t) = 2p^{(\psi^+)}(m_1,\cdots,m_M;t)$ and $p^{Fermions}(m_1,m_2,\cdots,m_M;t) = 2p^{(\psi^-)}(m_1,\cdots,m_M;t)$. Now we suppose that $m_1 = m_2 = \cdots = m_M$, then for bosons one has

$$p^{Bosons}(m_1,\cdots,m_1;t) = 4\left|\psi_{LL\cdots LL}{}^{LR\cdots LR}(m_1,m_1,\cdots,m_1;t)\right|^2 + 4\left|\psi_{RR\cdots RR}{}^{LR\cdots LR}(m_1,m_1,\cdots,m_1;t)\right|^2$$
$$+\left|\begin{matrix}\psi_{RR\cdots RR}{}^{LR\cdots LR}(m_1,m_1,\cdots,m_1;t)+\psi_{RL\cdots RL}{}^{LR\cdots LR}(m_1,m_1,\cdots,m_1;t)\\ +\psi_{LR\cdots RL}{}^{LR\cdots LR}(m_1,m_1,\cdots,m_1;t)+\psi_{RL\cdots LR}{}^{LR\cdots LR}(m_1,m_1,\cdots,m_1;t)\end{matrix}\right|^2$$

Similarly for fermions one has

$$p^{Fermions}(m_1,m_1,\cdots,m_1;t) = \left|\begin{matrix}\psi_{LR\cdots LR}{}^{LR\cdots LR}(m_1,m_1,\cdots,m_1;t)-\psi_{RL\cdots RL}{}^{LR\cdots LR}(m_1,m_1,\cdots,m_1;t)\\ -\psi_{LR\cdots RL}{}^{LR\cdots LR}(m_1,m_1,\cdots,m_1;t)-\psi_{RL\cdots LR}{}^{LR\cdots LR}(m_1,m_1,\cdots,m_1;t)\end{matrix}\right|^2$$

If we compare the last two expressions for those obtained in the case of the Bell-type states we also get for the case $m_1 = m_2 = \cdots = m_{M-1} = m_M$, the following

$p^{Bosons}(m_1,m_1,\cdots,m_1;t) = p^{\psi^+}(m_1,m_1,\cdots,m_1;t)$ and $p^{Fermions}(m_1,m_2,\cdots,m_1;t) = p^{\psi^-}(m_1,m_1,\cdots,m_1;t)$.

Recall for indistinguishable particles we had

$P_{sameside}(t) = \sum_{i=1}^{M}\left[\sum_{m_i+1=-t}^{0}\left(\sum_{m_i+1=m_i}^{0} p(m_1,m_2,\ldots,m_M,t)\right)+\sum_{m_{i+1}=1}^{t}\left(\sum_{m_i=m_i+1}^{t} p(m_1,m_2,\ldots,m_M,t)\right)\right]$, where the summation is restricted to ordered pairs $(m_1,m_2,\cdots,m_M)$ with $m_i \geq m_{i+1}$. However using the results

$p^{Bosons}(m_1,m_2,\cdots,m_M;t) = 2p^{(\psi^+)}(m_1,\cdots,m_M;t)$, $p^{Fermions}(m_1,m_2,\cdots,m_M;t) = 2p^{(\psi^-)}(m_1,\cdots,m_M;t)$,

$p^{Bosons}(m_1,m_1,\cdots,m_1;t) = p^{\psi^+}(m_1,m_1,\cdots,m_1;t)$ , $p^{Fermions}(m_1,m_2,\cdots,m_1;t) = p^{\psi^-}(m_1,m_1,\cdots,m_1;t)$, we can replace $p^{(Bosons,Fermions)}(m_1,m_2,\cdots,m_M;t)$ in

$P_{sameside}(t) = \sum_{i=1}^{M}\left[\sum_{m_i+1=-t}^{0}\left(\sum_{m_i+1=m_i}^{0} p(m_1,m_2,\ldots,m_M,t)\right)+\sum_{m_{i+1}=1}^{t}\left(\sum_{m_i=m_i+1}^{t} p(m_1,m_2,\ldots,m_M,t)\right)\right]$ by

$p^{\psi^\pm}(m_1,\cdots,m_M,t)$ and extend the summation over $m_1,\cdots,m_M$ to conclude that

$P_{sameside}^{Bosons}(t) = P_{samside}^{\psi^+}(t)$ and $P_{sameside}^{Fermions}(t) = P_{sameside}^{\psi^-}(t)$, which shows that results obtained are the same for distinguishable particles starting the quantum walk with entangled coin states.

**7. On the notion of $\delta-$interaction: Fourier Analysis of $C_\delta$**

We begin by defining the evolution operator for quantum walks with $\delta-$interaction.

To define this, we change the factorized time evolution on the position space of the composite system given by $U^{composite} = U_1 \otimes U_2 \otimes U_3 \otimes \ldots \otimes U_{M-1} \otimes U_M$. Recall that in the original time evolution the coin was the same factorized coin in all lattice points $(m_1, m_2, \cdots, m_M)$, in the $\delta-$interaction we change the coin to a non-factorized one $C_\delta$, when the particles are at the same lattice point $m_1 = m_2 = m_3 = \cdots = m_M$. We then define the unitary time evolution operator for the quantum walk with $M$ $\delta-$interacting particles on a line as

$U_\delta = S^{composite}\left(\overline{P}_\delta \otimes \left(\overbrace{U_C \otimes \cdots U_C}^{M-times}\right)\right) + S^{composite}(P_\delta \otimes C_\delta)$, where $P_\delta$ is the projector on the joint position state of the $M-$particles, and is defined by $P_\delta = \sum_{m_1}\left(\overbrace{|m_1\rangle|m_1\rangle\cdots|m_1\rangle}^{M-times}\right)\left(\overbrace{\langle m_1|\langle m_1|\cdots\langle m_1|}^{M-times}\right)$,

and $\overline{P}_\delta = I^{composite} - P_\delta$.

In the context of the previous paragraph it should be noted that $C_\delta$ has dimension $2^M \times 2^M$, however here we will restrict ourselves to the case $M = 2$, and consider $C_\delta = \frac{1}{2}\begin{pmatrix} 1 & 1 & 1 & 1 \\ 1 & -1 & -1 & 1 \\ -1 & 1 & -1 & 1 \\ -1 & -1 & 1 & 1 \end{pmatrix}$, this coin can be considered the unfactorized version of the Hadamard coin in two dimensions, in that it is similar to the Hadamard coin in two dimensions, however, $C_\delta \neq H^*\left(\frac{1}{2}, \frac{1}{2}\right) \otimes H^*\left(\frac{1}{2}, \frac{1}{2}\right)$. Morever, recall that the 2-particle walk on the line is equivalent to a single particle walk in two dimensions. Now the spatial

Fourier transform $\widehat{\Psi}(k_x, k_y, t)$ for $k_x, k_y \in [-\pi, \pi]$ of the wave function $\Psi(x, y, t)$ over $Z^2$ is given by $\widehat{\Psi}(k_x,k_y,t)=\sum_{(k_x,k_y)\in Z^2}\Psi(x,y,t)e^{i(k_x x+k_y y)}$. Let $\Psi(x,y,t)=\begin{pmatrix}\Psi_L(x,y,t)\\ \Psi_R(x,y,t)\\ \Psi_D(x,y,t)\\ \Psi_U(x,y,t)\end{pmatrix}$, where the subscripts

$L,R,D,$ and $U$ refer to the "left", "right", "down" and "up" chirality states, that is,

$\Psi(x,y,t)=\begin{pmatrix}\Psi_L(x,y,t)\\ \Psi_R(x,y,t)\\ \Psi_D(x,y,t)\\ \Psi_U(x,y,t)\end{pmatrix}$ is the four component vector of amplitudes of the particle being at point

$(x,y)$ at time $t$.

Define $M_-^L=\frac{1}{2}\begin{pmatrix}1&1&1&1\\0&0&0&0\\0&0&0&0\\0&0&0&0\end{pmatrix}$, $M_+^R=\frac{1}{2}\begin{pmatrix}0&0&0&0\\1&-1&-1&1\\0&0&0&0\\0&0&0&0\end{pmatrix}$, $M_-^D=\frac{1}{2}\begin{pmatrix}0&0&0&0\\0&0&0&0\\-1&1&-1&1\\0&0&0&0\end{pmatrix}$,

$M_+^U=\frac{1}{2}\begin{pmatrix}0&0&0&0\\0&0&0&0\\0&0&0&0\\-1&-1&1&1\end{pmatrix}$. The dynamics of the $C_\delta-$walk is given by

$\Psi(x,y,t+1)=M_+^R\Psi(x-1,y,t)+M_-^L\Psi(x+1,y,t)+M_+^U\Psi(x,y-1,t)+M_-^D\Psi(x,y+1,t)$. From

the dynamics of $\Psi$ we may deduce the following about $\widehat{\Psi}$:

$$\hat{\Psi}(k_x,k_y,t+1)=\sum_{x,y}\left[M_+^R\Psi(x-1,y,t)+M_-^L\Psi(x+1,y,t)+M_+^U\Psi(x,y-1,t)+M_-^D\Psi(x,y+1,t)\right]e^{i(k_x x+k_y y)}$$

$$=e^{ik_x}M_+^R\sum_{x,y}\Psi(x-1,y,t)e^{i(k_x(x-1)+k_y y)}+e^{-ik_x}M_-^L\sum_{x,y}\Psi(x+1,y,t)e^{i(k_x(x+1)+k_y y)}+e^{ik_y}M_+^U\sum_{x,y}\Psi(x,y-1,t)e^{i(k_x x+k_y(y-1))}+$$

$$e^{-ik_y}M_-^D\sum_{x,y}\Psi(x,y+1,t)e^{i(k_x x+k_y(y+1))}$$

$$=\left[e^{ik_x}M_+^R+e^{-ik_x}M_-^L+e^{ik_y}M_+^U+e^{-ik_y}M_-^D\right]\widehat{\Psi}(k_x,k_y,t)$$

This implies that we have $\hat{\Psi}(k_x,k_y,t+1)=M_{k_x,k_y}\hat{\Psi}(k_x,k_y,t)$, where

$$M_{k_x,k_y}=e^{ik_x}M_{+}^{R}+e^{-ik_x}M_{-}^{L}+e^{ik_y}M_{+}^{U}+e^{-ik_y}M_{-}^{D}=\frac{1}{2}\begin{pmatrix} e^{-ik_x} & e^{-ik_x} & e^{-ik_x} & e^{-ik_x} \\ e^{ik_x} & -e^{ik_x} & -e^{ik_x} & e^{ik_x} \\ -e^{ik_y} & e^{ik_y} & -e^{ik_y} & e^{ik_y} \\ -e^{-ik_y} & -e^{-ik_y} & e^{-ik_y} & e^{-ik_y} \end{pmatrix}$$

Note that we can write $M_{k_x,k_y}=D(e^{-ik_x},e^{ik_x},e^{ik_y},e^{-ik_y})\cdot C_\delta$, where $D(e^{-ik_x},e^{ik_x},e^{ik_y},e^{-ik_y})$ is the diagonal matrix with entries as shown and $C_\delta$ is the unitary matrix mentioned earlier that acts on the chirality state of the particle. Hence, $M_{k_x,k_y}$ is also a unitary matrix. In terms of the initial spatial Fourier transform we can write $\hat{\Psi}(k_x,k_y,t)=M^t_{k_x,k_y}\widehat{\Psi}(k_x,k_y,0)$. A popular method to calculate $M^t_{k_x,k_y}$ is to diagonalize the unitary matrix $M_{k_x,k_y}$, this method has been used successfully by a number of authors including Ampadu, Watabe et.al. Now we notice if $M_{k_x,k_y}$ has normalized eigenvectors $\left(N^1_{k_x,k_y}\left|\Phi^1_{k_x,k_y}\right\rangle, N^2_{k_x,k_y}\left|\Phi^2_{k_x,k_y}\right\rangle, N^3_{k_x,k_y}\left|\Phi^3_{k_x,k_y}\right\rangle, N^4_{k_x,k_y}\left|\Phi^4_{k_x,k_y}\right\rangle\right)$, where $N^i_{k_x,k_y}$ are appropriate normalization constants, and corresponding eigenvalues $\left(\lambda^1_{k_x,k_y},\lambda^2_{k_x,k_y},\lambda^3_{k_x,k_y},\lambda^4_{k_x,k_y}\right)$, then the evolution matrix $M^t_{k_x,k_y}$ can be written as $M^t_{k_x,k_y}=\sum_{i=1}^{4}\left(\lambda^i_{k_x,k_y}\right)^t\left(N^i_{k_x,k_y}\right)^2\left|\Phi^i_{k_x,k_y}\right\rangle\left\langle\Phi^i_{k_x,k_y}\right|$. It follows that the Fourier transform at time $t$ is given by

$$\widetilde{\Psi}(k_x,k_y,t)=\sum_{i=1}^{4}\left(\lambda^i_{k_x,k_y}\right)^t\left(N^i_{k_x,k_y}\right)^2\left\langle\Phi^i_{k_x,k_y}\middle|\widetilde{\Psi}(k_x,k_y,0)\right\rangle\left|\Phi^i_{k_x,k_y}\right\rangle.$$

**8. Concluding Remarks**

In this paper we have considered directional correlations between $M-$particles on a line. For non-interacting particles we have found analytic asymptotic expressions. When $\delta$-interaction is introduced in the model we have studied the Fourier analysis and obtain general analytic formula for the wave function of the walk in the case $M=2$ for the transformation $C_\delta$, which can be considered an unfactorized version of the Hadamard walk in two-dimensions. It is an interesting problem to deduce further properties of the transformation $C_\delta$ and other developed coins under the model of the quantum walk with $\delta-$interaction.